\documentclass[12pt]{report}
\usepackage{psfig}
\begin{document}
\title{  Prediction of Long Term Stability by Extrapolation }
\author{G. Parzen}
\date{ June 2000  \\BNL \\ C-A/AP/18 }
\maketitle
\tableofcontents

\begin{abstract}
\noindent
This paper studies the possibility of using the survival  function,
 to predict long term stability by extrapolation. The survival
function is a function of the initial coordinates and is the number of turns
a particle will survive for a given set of initial coordinates.
To determine the difficulties in extrapolating the survival function,
tracking studies were done to compute the survival  function. The survival 
 function was found to have two properties that may cause difficulties in
extrapolating the survival  function. One is the existence of rapid 
oscillations, and the second is the existence of plateaus. 
It was found that 
 it appears possible to extrapolate 
the survival  function to estimate long term stability by taking the two 
difficulties into account.     A model is proposed  which pictures the survival function to be a
series of plateaus  with rapid oscillations superimposed on the plateaus.
The tracking studies give results for the widths of these plateaus and for the 
seperation between adjacent plateaus which can be used to extrapolate and 
estimate the location of plateaus that indicate survival for longer times 
than can be found by tracking.
\end{abstract}

\pagestyle{headings}

\chapter{Introduction}
	This paper studies the possibility of using the survival  function,
 to predict long term stability by extrapolation ~\cite{k1,k2,k2b,k3,k4,k5,k5b}.
The survival
function is a function of the initial coordinates and is the number of turns
a particle will survive for a given set of initial coordinates.
To determine the difficulties in extrapolating the survival  function,
tracking studies were done to compute the survival  function. The survival 
function was found to have two properties that may cause difficulties in
extrapolating the survival turns function. One is the existence of rapid 
oscillations, and the second is the existence of plateaus.It was found that 
 it appears possible to extrapolate 
the survival  function to estimate long term stability by taking the two 
difficulties into account.

      A model is proposed  which pictures the survival function to be a
series of plateaus  with rapid oscillations superimposed on the plateaus.
The tracking studies give results for the widths of these plateaus and for the 
seperation between adjacent plateaus which can be used to extrapolate and 
estimate the location of plateaus that indicate survival for longer times 
than can be found by tracking.

\chapter {The survival  function, $j_{trns}$. }

     For a given set of initial coordinate, $x_0,p_{x0},y_0,p_{y0}$,  one can find  by tracking the survival time in turns, which is the number of turns the particle will survive before becoming unstable and which will be denoted by $j_{trns}$.  This determines the function $j_{trns}(x_0,p_{x0},y_0,p_{y0})$ which will be called the survival  function  ~\cite{k6,k7,k8,k8b}. If one limits the tracking to $1.0\; 10^6$ turns or less, one can find $j_{trns}$ for those $x_0,p_{x0},y_0,p_{y0}$ for which 
$j_{trns}$ is less than or equal to $1.0\; 10^6$.

	A tracking study  was done of particle motion   with no rf present,
using an older version of the RHIC lattice. Random and systematic field 
error multipoles are present up to order 10.
The particle momemtum is $dp/p=0$. As the first case studied, the initial coordinates $x_0,p_{x0},y_0,p_{y0}$ are chosen along the direction in phase space given by $p_{x0}=0,p_{y0}=0$ and $\epsilon_{x0}=\epsilon_{y0}$, where $\epsilon_{x0},\epsilon_{y0}$ are the linear emittances in the absence of the error multipoles.
Along this direction in phase space , $j_{trns}$ may be considered to be a function
of $x_0$. For a given  initial coordinate, $x_0$,  one can find  by tracking the survival time in turns, which is the number of turns the particle will survive before becoming unstable and will be denoted by $j_{trns}$.  This determines the survival function~\cite{k6,k7,k8,k8b} $j_{trns}(x_0)$.
 If one limits the tracking to $1\;10^6$ turns or less, one can find $j_{trns}$
for those $x_0$ for which $j_{trns}$ is less than or equal to $1\;10^6$. For $dp/p=0$, it is assumed that on the closed orbit $x=0$ and $x_0$ is also the initial betatron oscillation amplitude.

\begin{figure}[tbh]
\centerline{\psfig{file=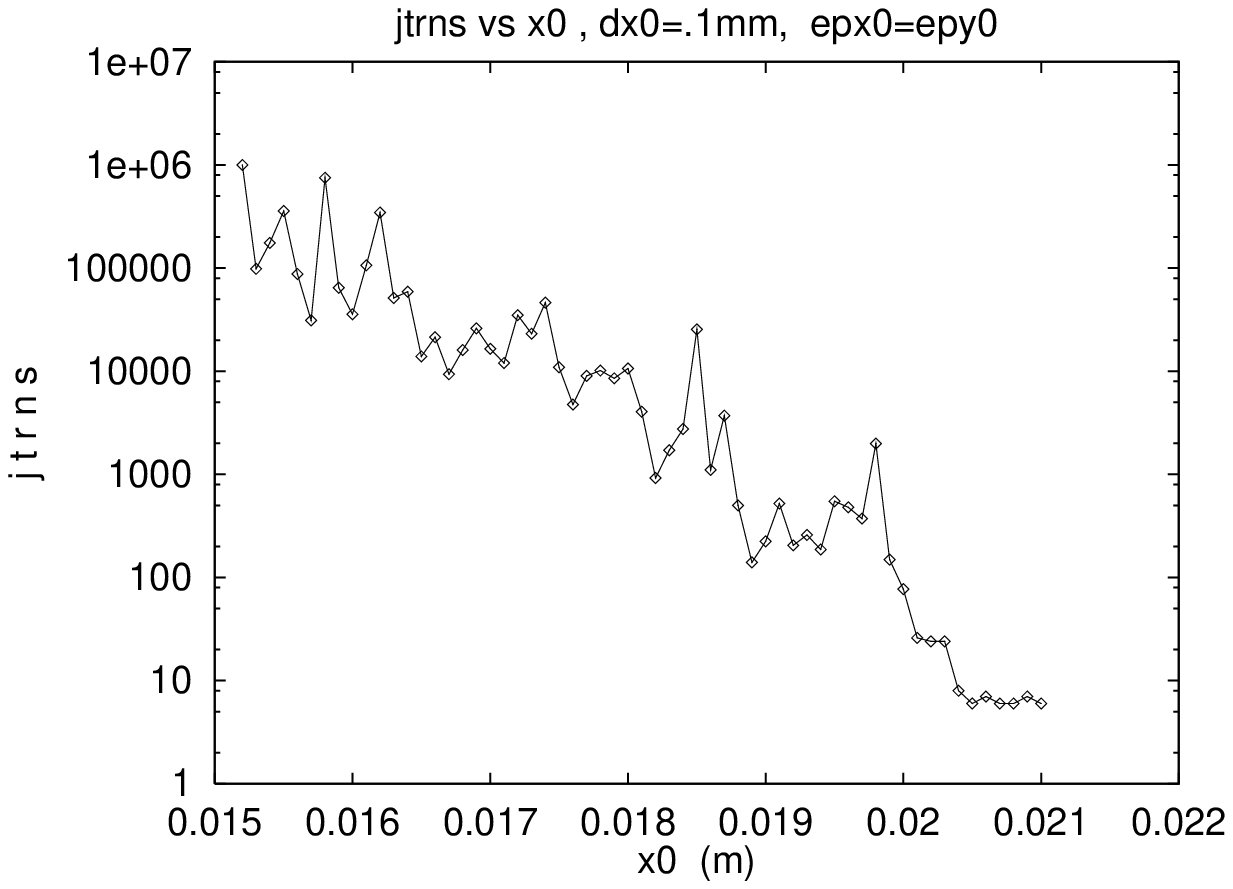,width=5.0in}}
\caption{$j_{trns}$ versus $x_0$. $dp/p=0$, $p_{x0}=0,p_{y0}=0,\epsilon_{x0}=\epsilon_{y0}$ direction, $dx_0$=.1mm. In the figure, jtrns, x0
, dx0, epx0, epy0  represent $j_{trns}$, $x_0$, $dx_0$ ,$\epsilon_{x0}$, $\epsilon_{y0}$. }
\label{fig.2-1}
\end{figure}

	The tracking was first done using $x_0$ which are seperated by $dx_0=.1$ mm. The results for $j_{trns}$ versus $x_0$ are shown in Fig.~\ref{fig.2-1}.
The results in this figure may be looked at  as the results of a search in $x_0$ starting at large $x_0$ and decreasing $x_0$ in steps of  $dx_0=.1 mm$ . The figure shows an apparent stability limit for 1e6 turns  of $u_{sl}=$15.2 mm.  

\section{Rapid oscillations in the survival function}

\begin{table}[tbh]
\begin{tabular}{|c|c|c|c|}\hline
$dx_0$ (mm)&$\Delta x_0$ (mm) &$u_{sl}$ (mm)&$j_{trns}$ at $u_{sl}+dx_0$\\ \hline
 .1	& .3	&15.2   & 100000	\\
 .01	& .06	&15.27  & 150000	\\
 .001	& .003	&15,270 & 82605		\\
 .0001	& .0002	&15.2793& 60693		\\  \hline
\end{tabular}
\caption{Results for different search intervals, $dx_0$.  $dx_0$ is decreased from .1mm to .0001 mm . $\Delta x_0$ is the wavelength of the oscilations , as measured from the stability limit for 1e6 turns, $u_{sl}$, to the location of the first peak in $j_{trns}$ . Also listed for each value of $dx_0$ are the apparent stability limit for 1e6 turns, $u_{sl}$, and $j_{trns}$ at $u_{sl}+dx_0$.}
\label{tab.2-1}
\end{table}

	Fig.~\ref{fig.2-1} shows rapid oscillations in $j_{trns}$ with $x_0$.
Large changes in $j_{trns}$ occur when $x_0$ changes by .1 mm. The oscillations
extend over about .3 mm. Tracking studies show that the oscillations become more
rapid when the search interval, $dx_0$, is decreased. This is indicated in 
Fig.~\ref{fig.2-2} where $dx_0$ is decreased to .05 mm and in Fig.~\ref{fig.2-3}
where $dx_0$ is decreased to .025 mm.
Results for different 
search intervals, $dx_0$, are shown in Table~\ref{tab.2-1}. $dx_0$ is decreased from .1mm to .0001 mm . The wavelength of the oscilations , $\Delta x_0$, as measured from the stability limit for 1e6 turns, $u_{sl}$, to the location of the first peak in $j_{trns}$ decreases from about .3 mm to .0002 mm. Also listed for each
value of $dx_0$ are the apparent stability limit for 1e6 turns, $u_{sl}$, and $j_{trns}$
at $u_{sl}+dx_0$.

	Table~\ref{tab.2-1} shows that the wavelength of the oscilations , $\Delta x_0$ is roughly proportional to the size of the search interval, $dx_0$.
The value of $j_{trns}$ at $u_{sl}+dx_0$ shows that near $u_{sl}$, $j_{trns}$
changes appreciably in the small change in $x_0$ given by $dx_0$. Tracking results
 show that this seems to hold even at extremely small $dx_0$.The computed results 
appear to indicate that $j_{trns}(x_0)$ is not a continuous function of $x_0$. For
a continuous function of $x_0$, one can find a small enough interval in $x_0$
such that the differnce between the values of the function for any two $x_0$ in
that interval is very small. This does not appear to be true for $j_{trns}(x_0)$.

	The existence of the rapid oscillations in the survival function,
$j_{trns}(x_0)$, would seem to make it difficult to extrapolate to find those
$x_0$ that survive for more than 1e6 turns. However one could view the survival
function shown in Fig.~\ref{fig.2-1} as being made up of the rapid oscillations 
superimposed on a smoother, more slowly varying function which could be used 
for the extrapolation. This is discussed further in section 2.3.

\section {Plateaus in the survival function}

\begin{figure}[tbh]
\centerline{\psfig{file=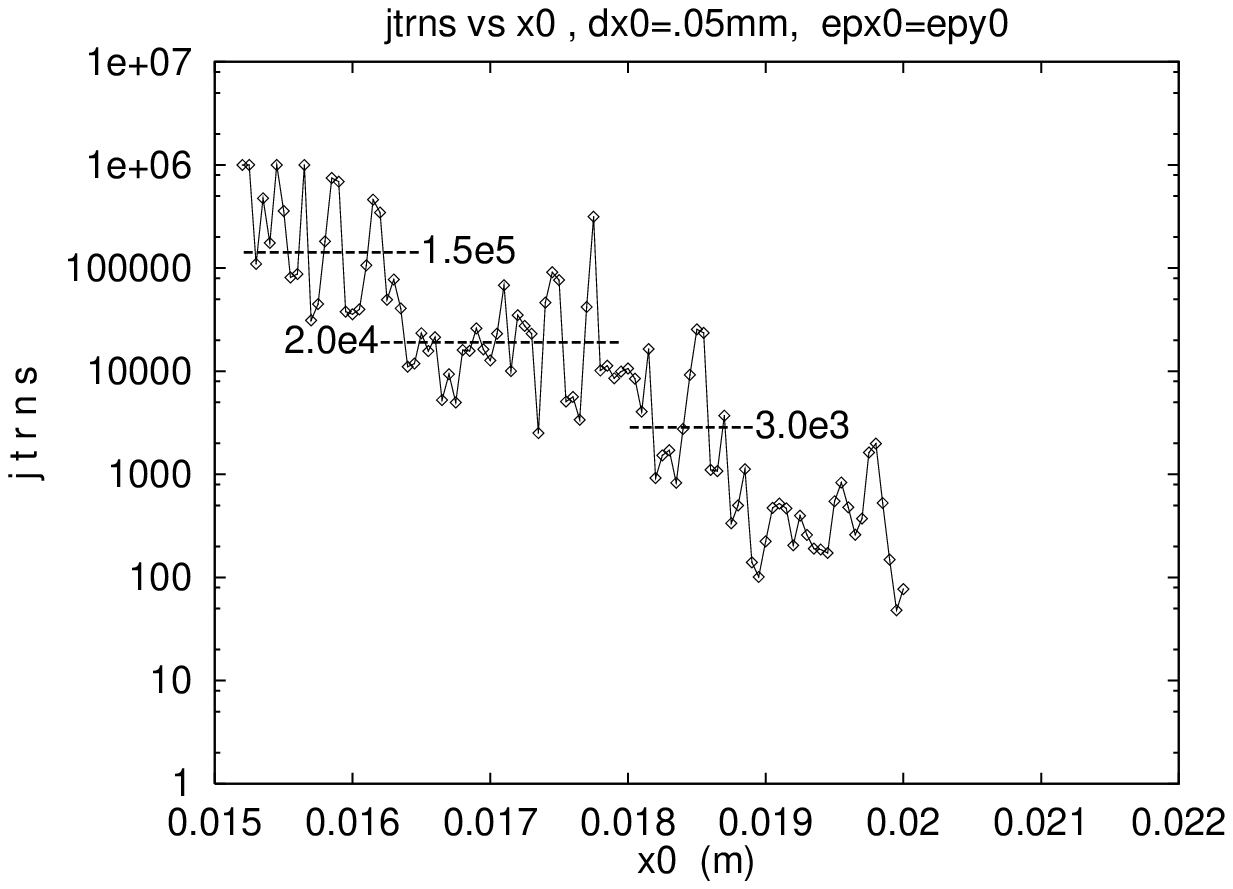,width=5.0in}}
\caption{$j_{trns}$ versus $x_0$. $dp/p=0$, $p_{x0}=0,p_{y0}=0,\epsilon_{x0}=\epsilon_{y0}$ direction, $dx_0$=.05mm. In the figure, jtrns, x0
, dx0, epx0, epy0  represent $j_{trns}$, $x_0$, $dx_0$ ,$\epsilon_{x0}$, $\epsilon_{y0}$. }
\label{fig.2-2}
\end{figure}

\begin{figure}[tbh]
\centerline{\psfig{file=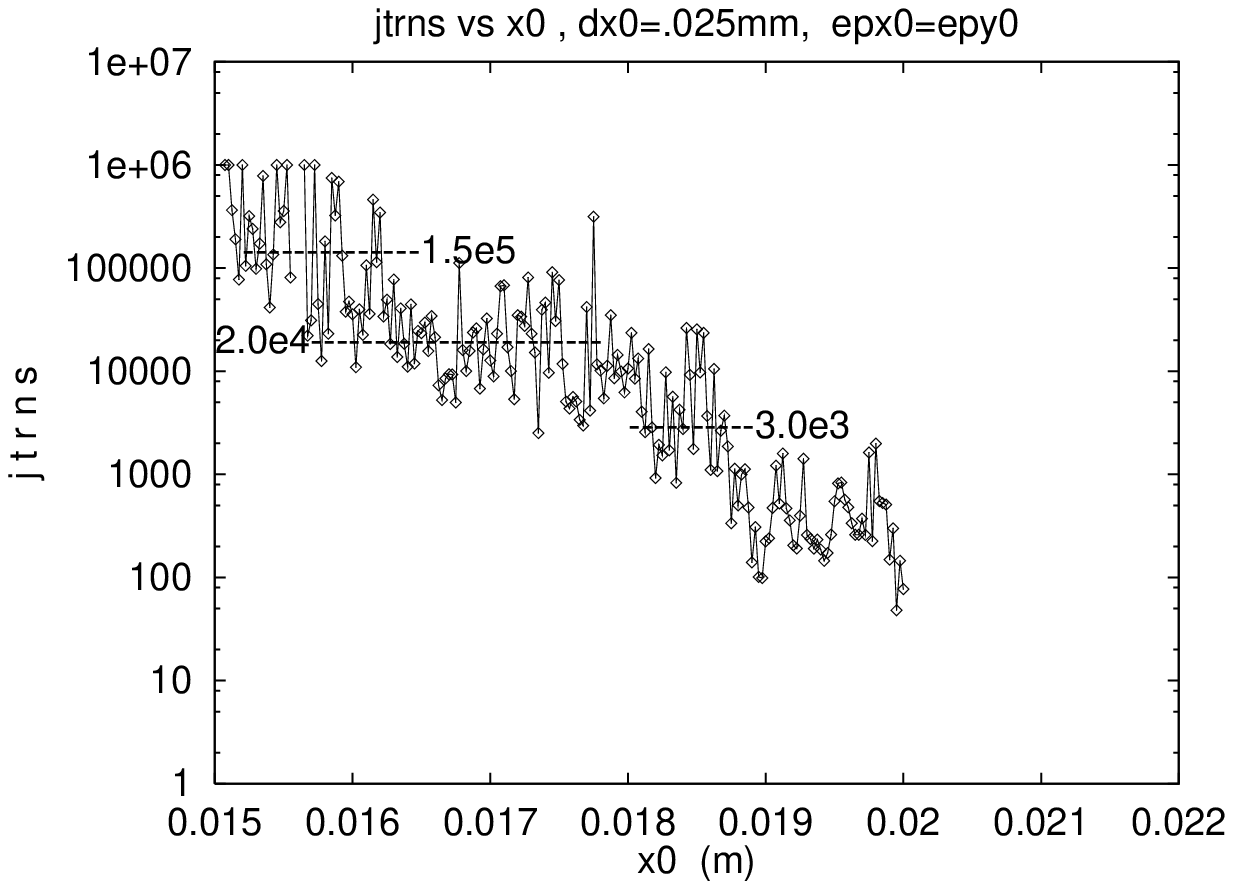,width=5.0in}}
\caption{$j_{trns}$ versus $x_0$. $dp/p=0$, $p_{x0}=0,p_{y0}=0,\epsilon_{x0}=\epsilon_{y0}$ direction, $dx_0$=.025mm. In the figure, jtrns, x0
, dx0, epx0, epy0  represent $j_{trns}$, $x_0$, $dx_0$ ,$\epsilon_{x0}$, $\epsilon_{y0}$. }
\label{fig.2-3}
\end{figure}

	Looking at Fig.~\ref{fig.2-1}, one can make out plateaus in the
survival function, $j_{trns}(x_0)$. The plateaus are regions where $j_{trns}$
oscillates rapidly around an almost constant value of $j_{trns}$. The plateaus
can be seen somewhat more clearly if one reduces the search interval $dx_0$,
as shown in Fig.~\ref{fig.2-2} where $dx_0$ is decreased to $dx_0=.05 mm$. One can make out 
4 plateaus located at about $j_{trns}$=1.5e5, 2e4, 3500, 400 turns. Possible
plateaus with $j_{trns}$ less than 100 turns are being ignored. It will be seen
that the width of the
plateaus do not depend on the search interval, $dx_0$. This is also true of the 
location of the plateaus in $j_{trns}$. This is  shown in Fig.~\ref{fig.2-3}
where $dx_0$=.025 mm. 

	The existence of the plateaus in the survival function,
$j_{trns}(x_0)$, would seem to make it difficult to extrapolate to find those
$x_0$ that survive for more than 1e6 turns. If one does the extrapolation
using points  which are close to the stability limit for 1e6 turns, $u_{sl}$,
one may get incorrect results as these points may lie on one of the plateaus.

\section{Extrapolation of the survival function}

\begin{figure}[tbh]
\centerline{\psfig{file=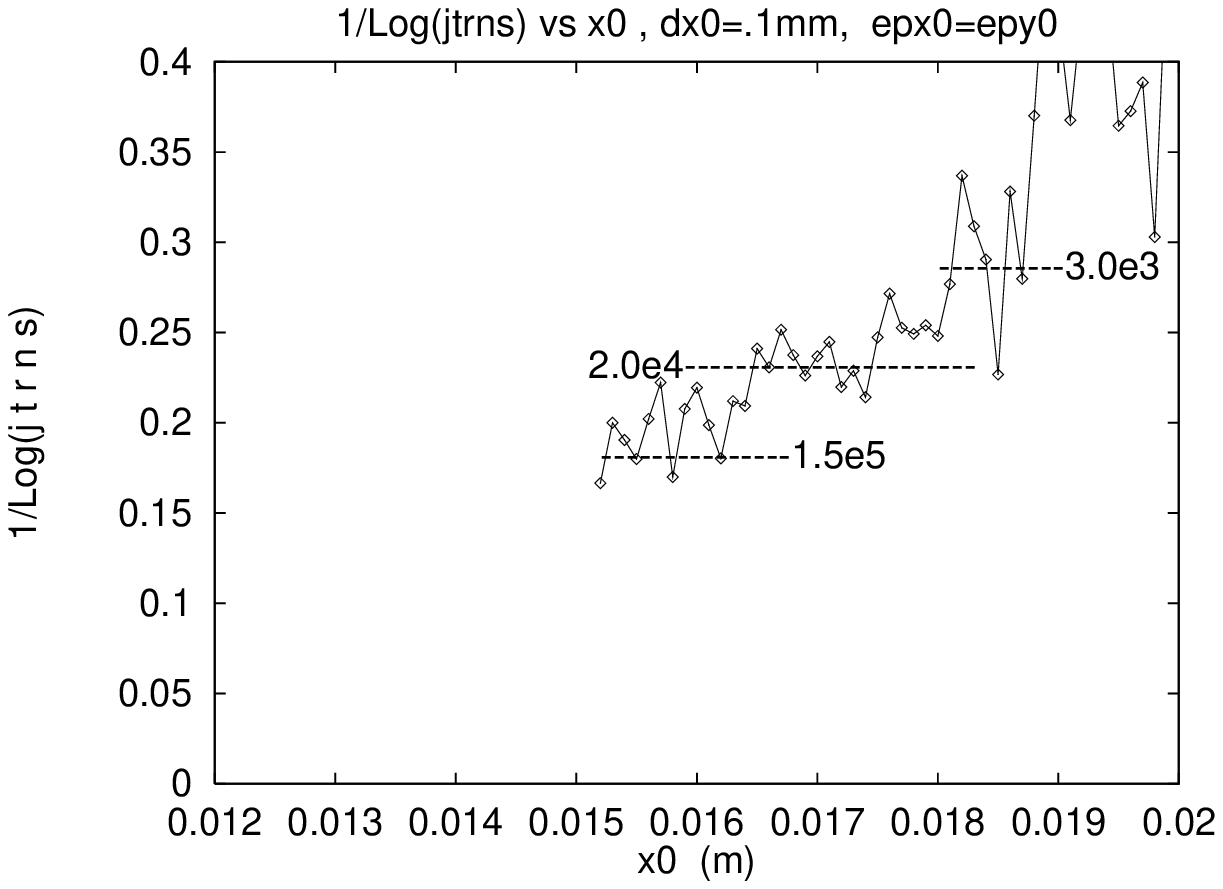,width=5.0in}}
\caption{$1/Log(j_{trns})$ versus $x_0$. $dp/p=0$, $p_{x0}=0,p_{y0}=0,\epsilon_{x0}=\epsilon_{y0}$ direction, $dx_0$=.1mm. In the figure, Log(jtrns), x0
, dx0, epx0, epy0  represent $1/Log(j_{trns})$, $x_0$, $dx_0$ ,$\epsilon_{x0}$, $\epsilon_{y0}$. }
\label{fig.2-4}
\end{figure}

\begin{figure}[tbh]
\centerline{\psfig{file=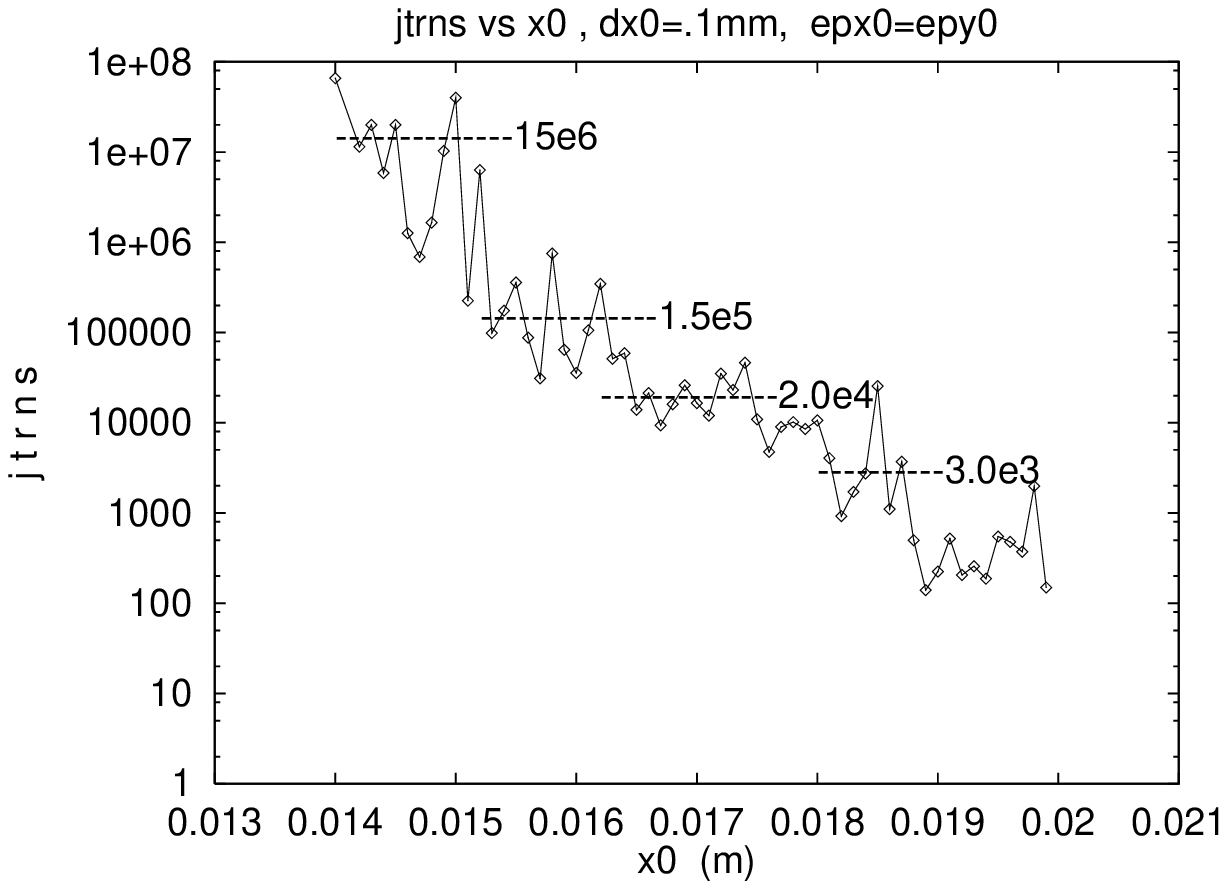,width=5.0in}}
\caption{$j_{trns}$ versus $x_0$ including points with $j_{trns}$ up to 4e7. $dp/p=0$, $p_{x0}=0,p_{y0}=0,\epsilon_{x0}=\epsilon_{y0}$ direction, $dx_0$=.1mm. In the figure, jtrns, x0
, dx0, epx0, epy0  represent $j_{trns}$, $x_0$, $dx_0$ ,$\epsilon_{x0}$, $\epsilon_{y0}$. }
\label{fig.2-5}
\end{figure}

\begin{figure}[tbh]
\centerline{\psfig{file=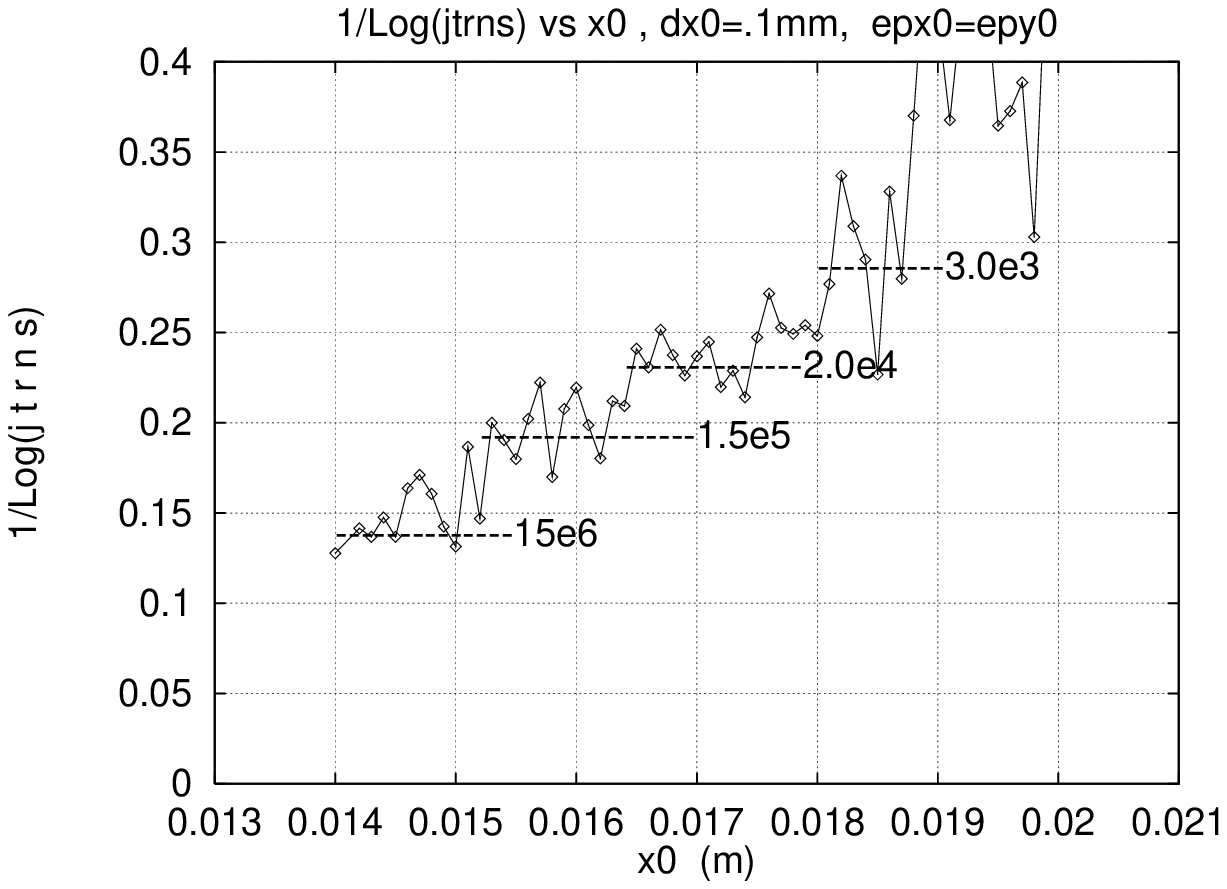,width=5.0in}}
\caption{$1/Log(j_{trns})$ versus $x_0$ showing the next plateau and the last 
measureed plateau. $dp/p=0$, $p_{x0}=0,p_{y0}=0,\epsilon_{x0}=\epsilon_{y0}$ 
direction, $dx_0$=.1mm. In the figure, 1/Log( jtrns), x0
, dx0, epx0, epy0  represent $1/Log(j_{trns})$, $x_0$, $dx_0$ ,$\epsilon_{x0}$, $\epsilon_{y0}$. }
\label{fig.2-6}
\end{figure}

\begin{figure}[tbh]
\centerline{\psfig{file=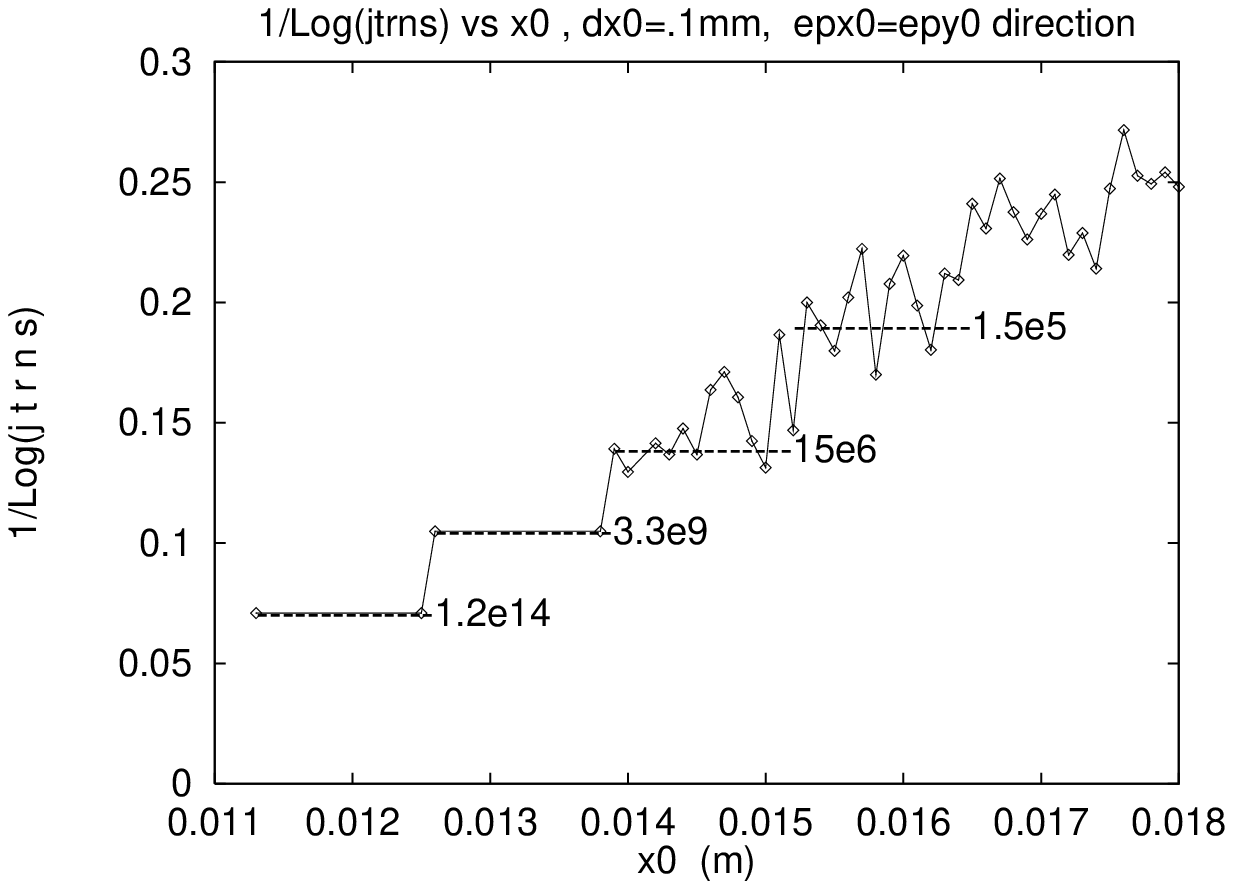,width=5.0in}}
\caption{$1/Log(j_{trns})$ versus $x_0$ showing the  plateaus found by
extrapolation, including the 1e9 plateau.
 $dp/p=0$, $p_{x0}=0,p_{y0}=0,\epsilon_{x0}=\epsilon_{y0}$ 
direction, $dx_0$=.1mm. In the figure, 1/Log( jtrns), x0
, dx0, epx0, epy0  represent $1/Log(j_{trns})$, $x_0$, $dx_0$ ,$\epsilon_{x0}$, $\epsilon_{y0}$. }
\label{fig.2-7}
\end{figure}

     The data given above leads to a model of the survival functioon, which
pictures it as sequence of plateaus. Within the plateaus, $j_{trns}$ oscillates about 
some constant vaue of $j_{trns}$ which will be called the level of the plateau.
The existence of the plateaus makes extrapolation of the survival function
appear difficult. The last plateau that was measured has a level of about 1.5e5 turns and a width of about 1.2 mm. An interesting question is what is the level of the next plateau, at lower $x_0$, after the last plateau that was measured. In the following this plateau will be referred to as the
'next plateau'.

	Some help with the extrapolation is also provided by plotting 
$1/Log(j_{trns})$ against $x_0$ as shown in Fig.~\ref{fig.2-4}, 
where the search interval is 
$dx_0$=.1mm and points  with $j_{trns}$  less than 400 turns are omitted. It will be seen below that the seperation between adjacent plateaus does not 
vary greatly when measured as the change in $1/Log(j_{trns})$. 

	To help locate the 'next plateau', long runs were done  starting 
with $x_0$=15.2mm, and decreasing $x_0$ 1n steps of .1mm. In order to detect 
the beginning of the 'next plateau', the runs have to be long enough not 
to be confused by the rapid oscillations in $j_{trns}$ that occur within each
plateau. Runs of length 2e7 turns were chosen, and these runs take about 10 days
for the RHIC lattice used. 
  
The results are shown in Fig.~\ref{fig.2-5} . One sees from
this figure that the 'next plateau' appears to start at $x_0$=15.3 mm and ends at about 14.0 mm, and has a level
of about $j_{trns}$=15e6 turns and a width of 1.2mm . The previous plateau goes from 16.5mm to 15.2 mm
and has a width of 1.3mm and a level of jtrns=1.5e5 turns. The width of the 
plateau is measured here from the end of one plateau in x0 to the end
of the adjacent plateau and includes the transition region where the points move
from one plateau to the next. The width of the 'next plateau' is difficult to
measure, as the adjacent plateau at still lower $x_0$ is estimated to have a 
level of 3e9 turns, and cannot be found by tracking. One can see that the width is
larger than 1.2mm. The $x_0$ at 14.0 mm survived
more than 89.9e6 turns. The data given above for the 'next plateau' is somewhat
in error but it seems better to use it than to throw away this information. 

\begin{table}[tbh]
\begin{tabular}{|c|c|c|c|}\hline
plateau level, $j_{trns}$              & 15e6  &1.5e5  &2e4        \\
plateau level,$1/Log(j_{trns})$        &.139  &.193   &.232        \\
plateal seperation in $1/Log(j_{trns})$ & .054 &.039   &---         \\
plateau width, $\Delta x_0$ (mm)       & 1.2+  &1.3    &1.7         \\  \hline
\end{tabular}
\caption{Plateau parameters for the epx0=epy0 direction.}
\label{tab.2-2}
\end{table}

The results are 
also shown as a  $1/Log(j_{trns})$ versus $x_0$ plot in Fig.~\ref{fig.2-6}. One can 
see three plateaus  near the stability boundary at the levels of about 
jtrns= 15e6, 1.5e5,and 2e4 turns. The properties of the plateaus are summarized in
Table~\ref{tab.2-2}.
The seperation between the plateaus in the $1/Log(j_{trns})$ plot are given by
.054 and .039, so that the seperation between the plateaus in the $1/Log(j_{trns})$
are not too different. The widths of the  plateaus that were measured were
1.2+, 1.3 and 1.7 mm. 

      The plateau model    will now be used to 
extrapolate and investigate long term stability in RHIC. In RHIC, $j_{trns}$=1e9 turns
coresponds to a survival time of about 3.5 hours. The data given above will be 
used to extrapolate to find the plateau whose level is greater than or equal to
$j_{trns}$=1e9 turns. This plateau will be called the 1e9 plateau. 
The existence of the plateaus indicates that there are limits on the accuracy
one can hope to achieve by extrapolation. The two parameters one needs to extrapolate the survival function are the plateau width and the plateau level seperation.
One cannot be certain what these  parameters will be at $j_{trns}$ of the order of
1e9 turns. However, one can use the data found for these two parameters at
$j_{trns}$ of the order of 1e6 turns , to make the best estimate for these parameters
at $j_{trns}$ of the order of 1e9 turns.

The plateau width and the plateau level seperation were studied for three different cases coresponding to three different 
directions in phase space. The results are summarized in chapter 5.
The widths of the plateaus  were found to be roughly constant when measured in terms of $X_0$=$[\beta_{x0} (\epsilon_{x0} +\epsilon_{y0} )]^{.5}$ with an average  value $\Delta X_0$=2.00 mm.
 The plateau level seperation measured in $1/Log(j_{trns})$ varied from
.054 to .019 with an average value of .033. It is suggested that in the extrapolation, the average value of these two parameters be used.
For the case being considered here, it is assumed in the extrapolation that the plateau widths in $x_0$ will be 1.4 mm corresponding to the average value of   $\Delta X_0$=2.00 mm,  and the plateau level seperation is .033 in 
$1/Log(j_{trns})$. Note that in this case $X_0$=1.414 $x_0$
 
The results 
found using these assumptions are shown in Fig.~\ref{fig.2-7}, where two new plateaus
are shown that were found by extrapolation. The plateau just below the next 
plateau in x0 has the level of $j_{trns}$=3.3e9 turns and is the 1e9 plateau 
in this example. The 1e9 plateau goes from $x_0$=13.9mm to $x_0$=12.6mm. 
The aperture for 1e9 turns may be taken as about 13.9mm,
which may be compared to the 15.2 mm found using runs of 1e6 turns. This indicates
a loss of about 1.3mm or about 9\% due to the required survival time of 1e9 turns.
One may note that the level of the 1e9 plateau being 3.3e9 turns, one may
expect that some of the $x_0$ on this plateau will not survive 1e9 turns due to the oscillations in jtrns that will occur on this plateau.To be safer one could assume
the aperture for 1e9 turns to be 12.6 mm, which is the beginning of the adjacent
plateau with a level of 1.2e14 turns, giving a loss of 17\%.
 
The procedure
used indicates that in this case the result is not sensitive to changes in the 
two assumptions made regarding the width and level seperation of the plateaus.
If the plateau model continues to hold for the extrapolated plateaus, and the 
width and level seperation remain  roughly 1.3 mm and .034 respectively, then the 
result for the aperture for 1e9 turns will be about the same.

	The plateau model developed above avoids certain problems that arise
in tracking studies. In trying to find the aperture for the survival time of 1e9
turns, the more usual approach is to try find the $x_0$ such that all smaller
$x_0$ will survive for 1e9 turns. If while doing a tracking search starting from
large $x_0$, one finds a $x_0$ that survives for 1e9 turn, then one has to ask whether
all smaller $x_0$ will survive for 1e9 turns. This is a difficult question to answer.
It is even possible, that there are are always 
smaller $x_0$ that will not survive 1e9 turns, although these $x_0$ may become very
scarce at smaller $x_0$. In the plateau model, in trying to find the aperture 
for the survival time of 1e9 turns, the approach is to try to find the plateau
whose level is greater than  1e9 turns. This is a better defined 
target, and the expectation is that this plateau will indicate a region of $x_0$
where most $x_0$ will survive 1e9 turns.

\chapter{Survival function along the $\epsilon _{y0}=0$ direction}

\begin{figure}[tbh]
\centerline{\psfig{file=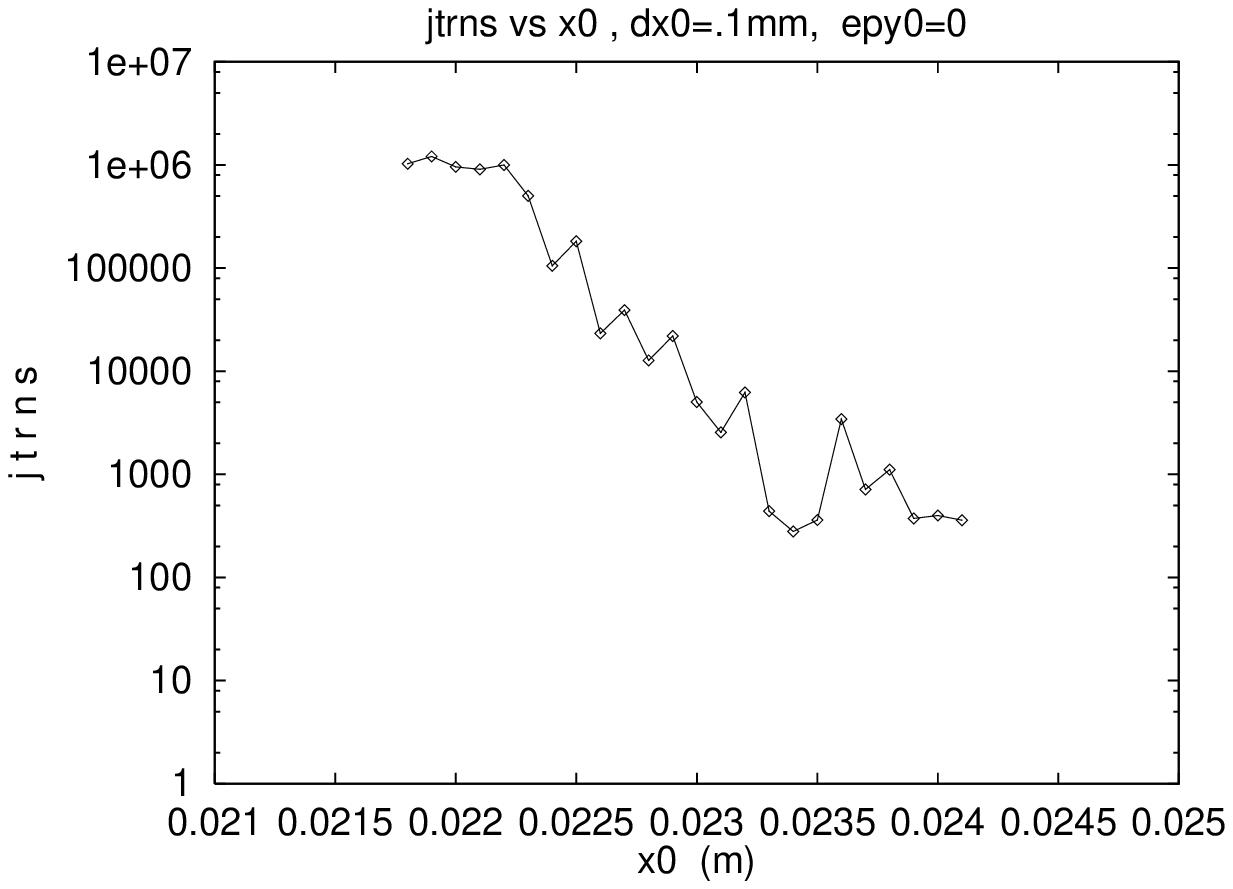,width=5.0in}}
\caption{$j_{trns}$ versus $x_0$ . $dp/p=0$, $p_{x0}=0,p_{y0}=0,\epsilon_{y0}=0$ 
direction, $dx_0$=.1mm. In the figure, jtrns, x0
, dx0, epx0, epy0  represent $j_{trns}$, $x_0$, $dx_0$ ,$\epsilon_{x0}$, $\epsilon_{y0}$. }
\label{fig.3-1}
\end{figure}

\begin{figure}[tbh]
\centerline{\psfig{file=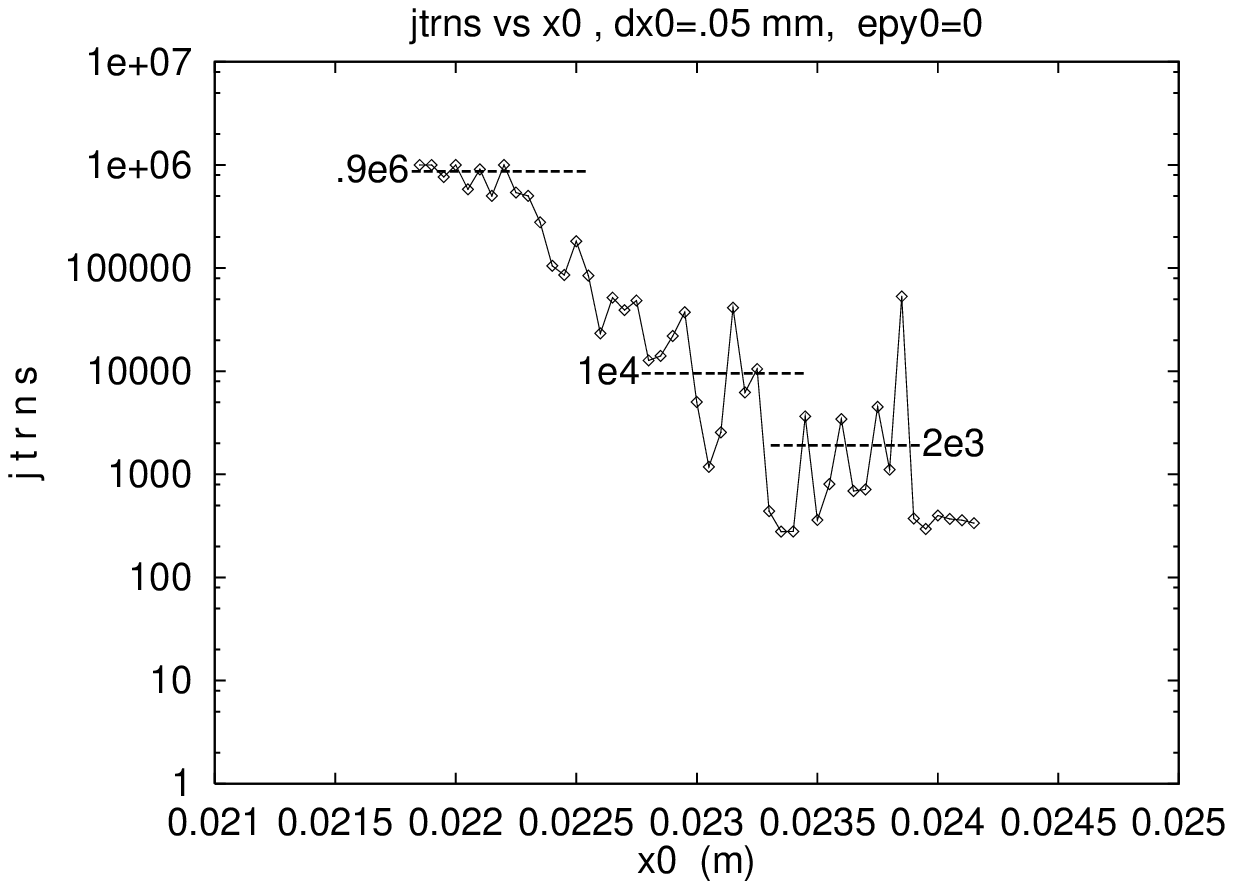,width=5.0in}}
\caption{$j_{trns}$ versus $x_0$ . $dp/p=0$, $p_{x0}=0,p_{y0}=0,\epsilon_{y0}=0$ 
direction, $dx_0$=.05mm. In the figure, jtrns, x0
, dx0, epx0, epy0  represent $j_{trns}$, $x_0$, $dx_0$ ,$\epsilon_{x0}$, $\epsilon_{y0}$. }
\label{fig.3-2}
\end{figure}

\begin{figure}[tbh]
\centerline{\psfig{file=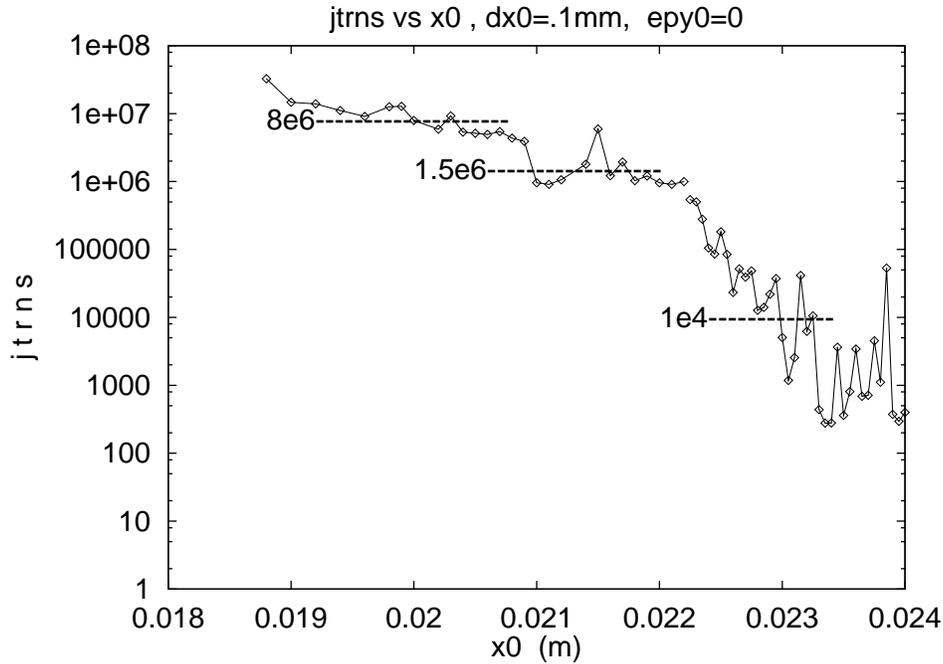,width=5.0in}}
\caption{$j_{trns}$ versus $x_0$ including points with $j_{trns}$ up to 4e7. $dp/p=0$, $p_{x0}=0,p_{y0}=0,\epsilon_{y0}=0$ 
direction, $dx_0$=.1 mm with $dx_0$=.05 mm at larger $x_0$. In the figure, 
jtrns, x0
, dx0, epx0, epy0  represent $j_{trns}$, $x_0$, $dx_0$ ,$\epsilon_{x0}$, $\epsilon_{y0}$. }
\label{fig.3-3}
\end{figure}

\begin{figure}[tbh]
\centerline{\psfig{file=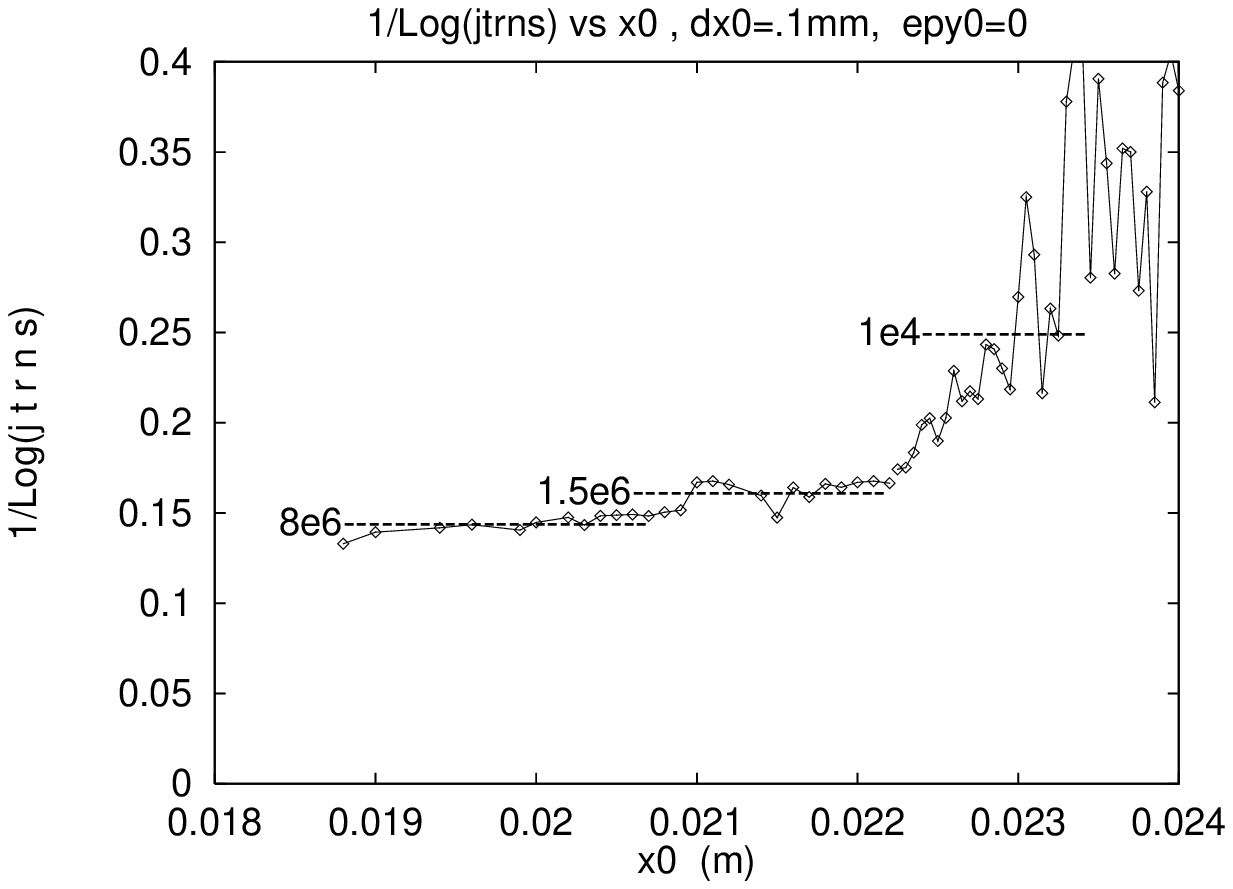,width=5.0in}}
\caption{$1/Log(j_{trns})$ versus $x_0$ . $dp/p=0$, $p_{x0}=0,p_{y0}=0,\epsilon_{y0}=0$ 
direction, $dx_0$=.1 mm. In the figure, Log( jtrns), x0
, dx0, epx0, epy0  represent $1/Log(j_{trns})$, $x_0$, $dx_0$ ,$\epsilon_{x0}$, $\epsilon_{y0}$. }
\label{fig.3-4}
\end{figure}

\begin{figure}[tbh]
\centerline{\psfig{file=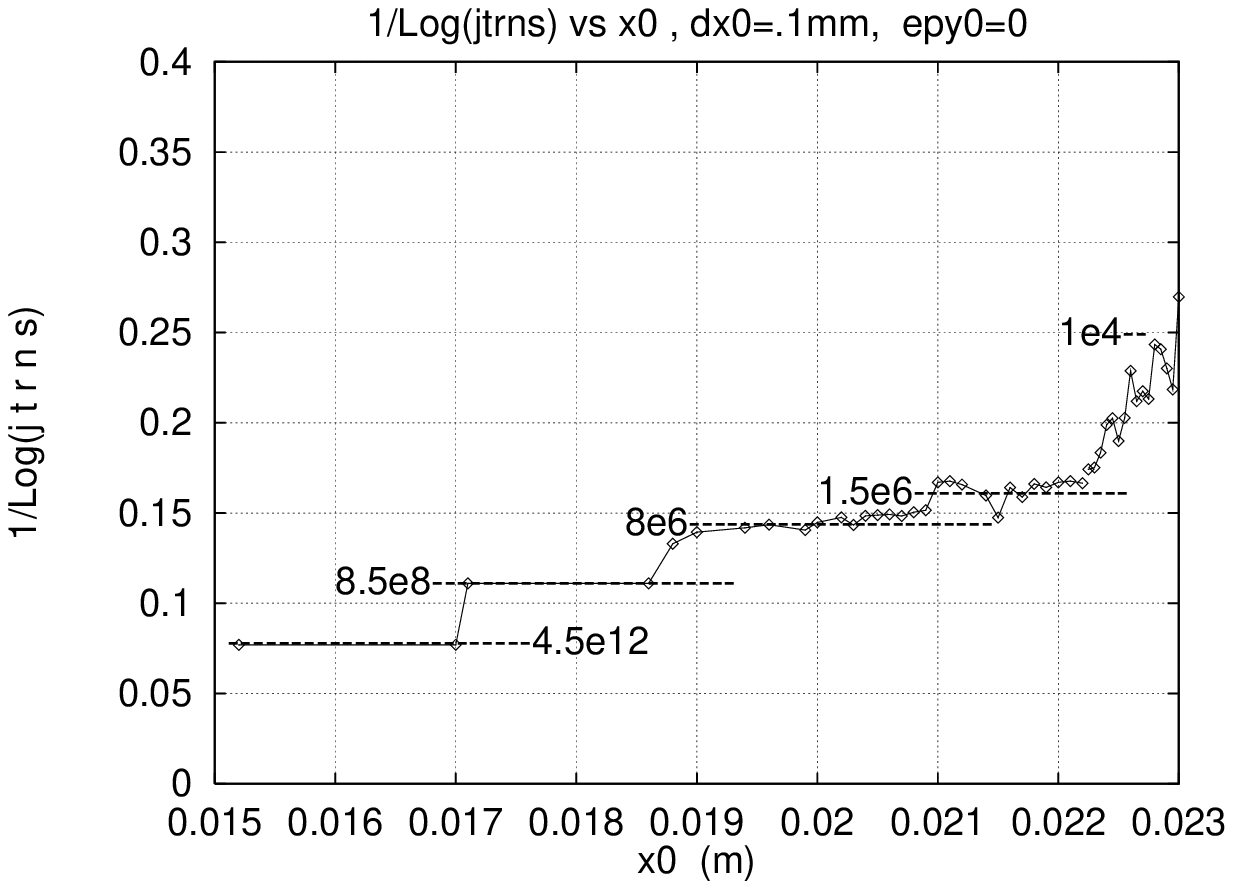,width=5.0in}}
\caption{$1/Log(j_{trns})$ versus $x_0$ showing the  plateaus found by
extrapolation, including the 1e9 plateau.
 $dp/p=0$, $p_{x0}=0,p_{y0}=0,\epsilon_{x0}=\epsilon_{y0}$ 
direction, $dx_0$=.1mm. In the figure, 1/Log( jtrns), x0
, dx0, epx0, epy0  represent $1/Log(j_{trns})$, $x_0$, $dx_0$ ,$\epsilon_{x0}$, $\epsilon_{y0}$. }
\label{fig.3-5}
\end{figure}

	The tracking results presented above were all done along the 
$\epsilon_{x0}=\epsilon_{y0}$ direction. Results will now be given for the $y_0=0,p_{y0}=0$ or  $\epsilon_{y0}=0$ direction.
Tracking studies of other cases will test the consistancy of the plateau model. The direction in the space of $x_0,p_{x0},y_0,p_{y0}$ is further defined by $p_{x0}=0$. 
The particle motion is 4 dimensional because of the presence of skew multipoles.
$j_{trns}$ may then be considered to be a function of $x_0$. Fig.~\ref{fig.3-1} shows $j_{trns}$ as a function 
of $x_0$ as found with tracking runs of 1e6 turns. The 
apparent stability limit using 1e6 turns and $dx_0$=.1mm is 22.0 mm. To make 
the plateaus more visible , one can reduce the search interval $dx_0$ to $dx_0$=.05 mm.
These results are shown in Fig.~\ref{fig.3-2}. One sees that there are 
two  plateaus in the region shown with $j_{trns}$ greater than or equal to 1e4,
 whose levels are located at $j_{trns}$= .9e6 ,
 and 1e4  turns .The  oscillations on the  plateau near the stability boundary,
 appear to be smaller than those seen in 
the $\epsilon_{x0}=\epsilon_{y0}$ case.  

Runs of about
2e7 turns were done to find the shape of the 'next plateau' in the 
region where $x_0$ is less than or equal to 21.0 mm. Enough tracking runs were done to determine
the beginning , and the level of the 'next plateau'. The results are shown in
Fig.~\ref{fig.3-3}. Including these longer runs , one sees that the 'next plateau'
begins at x0=21.0mm and the level of the 'next plateau' is about $j_{trns}$=8e6 turns.
Here, the begining of the plateau is the edge at larger $x_0$ and the end is the
edge at lower $x_0$.
Th end of the 'next plateau' is some what difficult to determine, as the 
adjacent plateau at lower $x_0$ has a high level of about 1e9 turns. 
The end of the ' next plateau' was taken to be at $x_0$=19.0 mm. 
The adjacent plateau at higher x0 is at the level of 1.5e6 turns 
and with the width of 1.8 mm. These two plateaus are seperated by .017 in 
$1/Log(j_{trns})$ which is smaller than the .054 found in the $\epsilon_{x0}$=$\epsilon_{y0}$ case. The 
results are also shown as $1/Log(j_{trns})$ versus $x_0$ plot in Fig.~\ref{fig.3-4}.

   The data found for these two plateaus will be used to extrapolate and find
the 1e9 plateau in this case. In the extrapolation, the results found in 
chapter 5 for the average plateau width and level seperation  will be used.
In this case, this leads to the assumptions 
 that each
plateau is 2.0mm wide and 
a plateau level seperation 
of .033 in $1/Log(j_{trns})$ will be used here.

 The 
results are shown in Fig.~\ref{fig.3-5}, where two  extrapolated plateaus 
are shown with the plateau levels 8.5e8, and 4.5e12 .  The aperture for 
1e9 turns was taken to be 17.1mm, the beginning of the plateau with a level
of 4.5e12 turns. 17.1 mm 
 is to be compared with the aperture of 21.9 mm found with runs 
of 1e6 turns. A loss of 4.8 mm or 22\%.

\chapter{ Survival function along  $\epsilon _{x0}=0$ direction}
\begin{figure}[tbh]
\centerline{\psfig{file=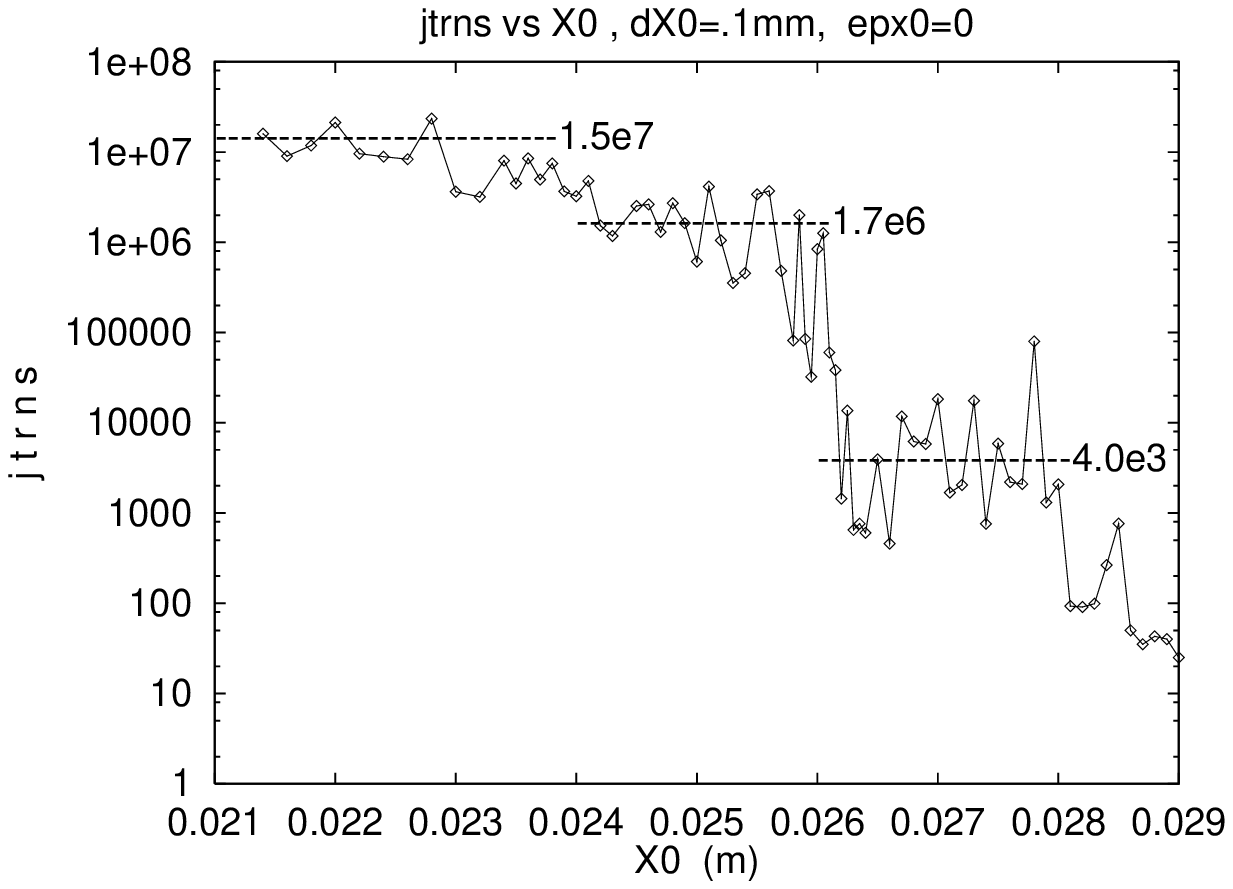,width=5.0in}}
\caption{$j_{trns}$ versus $X_0$ .$X_0=(\beta_{x0}/\beta_{y0})^{.5} y_0$. $dp/p=0$, $x_0$=0, 
$p_{x0}=0,p_{y0}=0, \epsilon_{x0}=0$ 
direction, $dx_0$=.1 mm  and runs go up to 2e7 turns.  In the figure, 
jtrns, X0
, dX0, epx0, epy0  represent $j_{trns}$, $X_0$, $dX_0$ ,$\epsilon_{x0}$, 
$\epsilon_{y0}$. }
\label{fig.4-1}
\end{figure}

\begin{figure}[tbh]
\centerline{\psfig{file=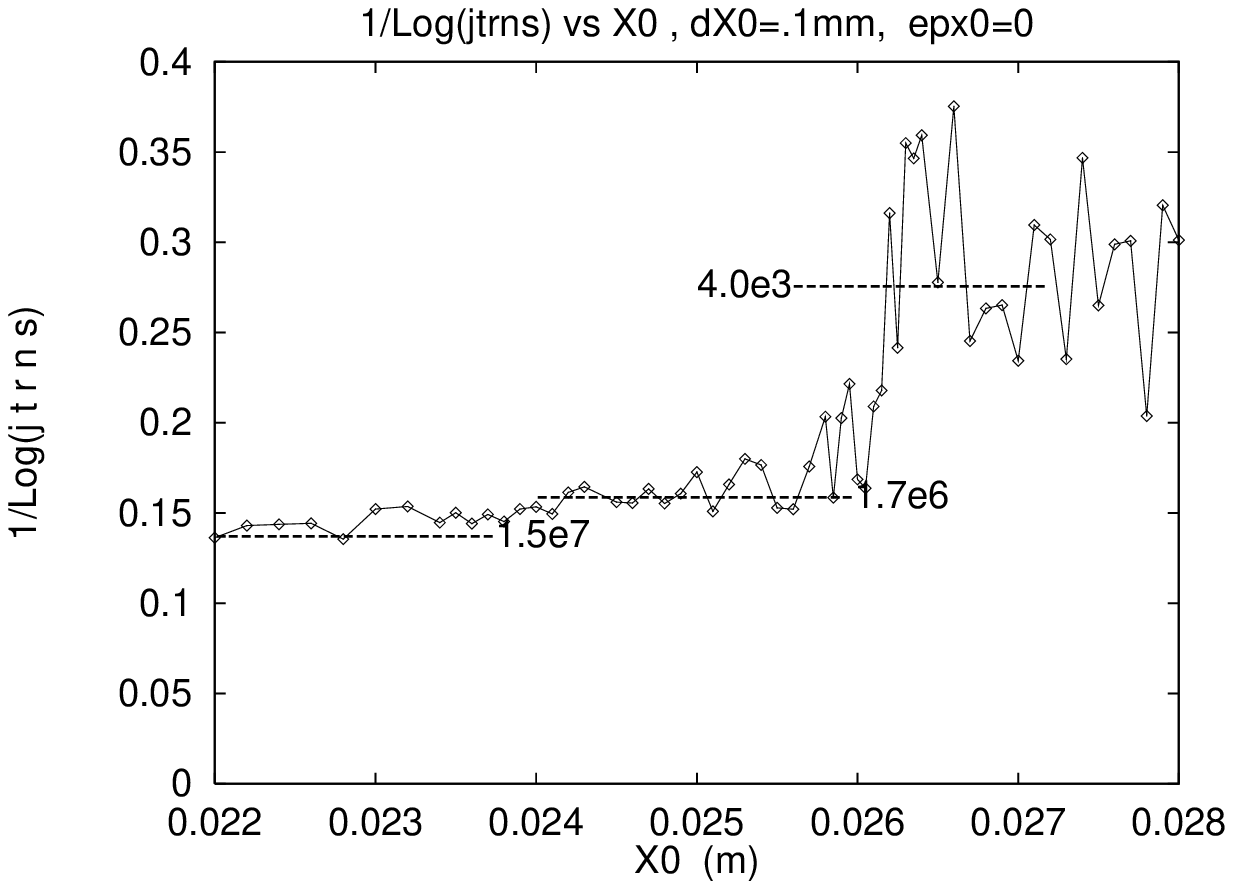,width=5.0in}}
\caption{$1/Log(j_{trns})$ versus $X_0$ . $dp/p=0$, $p_{x0}=0,p_{y0}=0,\epsilon_{x0}=0$ 
direction, $dX_0$=.1 mm. $X_0=(bex0/bey0)^{.5} y_0$. In the figure, 1/ Log( jtrns), X0
, dX0, epx0, epy0  represent $1/Log(j_{trns})$, $X_0$, $dX_0$ ,$\epsilon_{x0}$, $\epsilon_{y0}$. }
\label{fig.4-2}
\end{figure}

\begin{figure}[tbh]
\centerline{\psfig{file=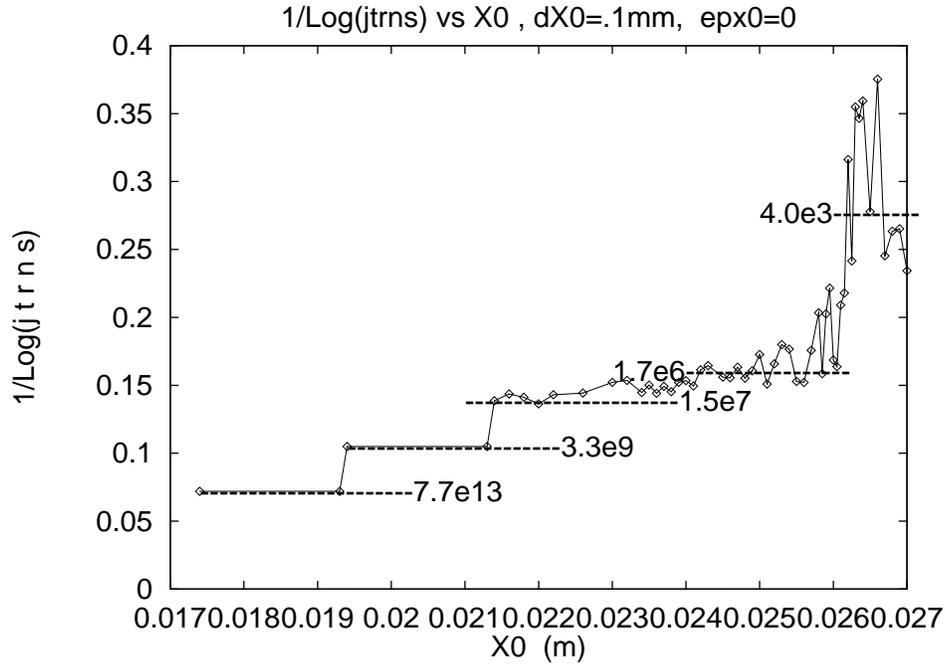,width=5.0in}}
\caption{$1/Log(j_{trns})$ versus $X_0$ . $dp/p=0$, $p_{x0}=0,p_{y0}=0,\epsilon_{x0}=0$ direction, $dX_0$=.1 mm, $X_0=(bex0/bey0)^{.5} y_0$, showing
 extrapolated plateaus.  
In the figure, 1/ Log( jtrns), X0
, dX0, epx0, epy0  represent $1/Log(j_{trns})$, $X_0$, $dX_0$ ,$\epsilon_{x0}$, $\epsilon_{y0}$. }
\label{fig.4-3}
\end{figure}

Results will now be given for the $x_0=0,p_{x0}=0$ or  $\epsilon_{x0}=0$ 
direction.In order to be able to compare results with those of the above 
two cases, $j_{trns}$ will plotted against $X_0=(\beta_{x0}/\beta_{y0})^{.5} y_0$.
Fig.~\ref{fig.4-1} shows $j_{trns}$ as a function of $X_0$ as found with 
tracking runs of about 2e7 turns.The apparent stability limit using 1e6 turns 
and $dx_0$=.1mm is $u_{sl}$=24.9 mm. In Fig.~\ref{fig.4-1} one can make out 
two plateaus with levels larger than 1e4 turns. The levels of these two 
plateaus are at 1.7e6 turns and 1.5e7 turns. The end of the 1.5e7 plateau
was taken as $X_0$=21.4 mm. The results are also shown as  a $1/Log(j_{trns})$ plot in 
Fig.~\ref{fig.4-2}.

   The data found for these two plateaus will be used to extrapolate and find
the 1e9 plateau in this case. In the extrapolation, the results found in 
chapter 5 for the average plateau width and level seperation  will be used.
In this case, this leads to the assumptions 
 that each
plateau is 2.0 mm wide and 
a plateau level seperation 
of .033 in $1/Log(j_{trns})$ will be used here.

 The 
results are shown in Fig.~\ref{fig.4-3}, where two  extrapolated plateaus 
are shown with the plateau levels 3.3e9, and 7.7e13 turns . 
Most of the particles on the 3.3e9 plateau will be assumed to survive 1e9 
turns, and the aperture for 1e9 turns is then 21.4mm.
21.4mm  is to be compared with the aperture of 24.9 mm found with runs 
of 1e6 turns. A loss of 3.5  mm or 17\%.

\chapter{ Extrapolation parameters for  the plateau model}

\begin{table}[tbh]
\begin{tabular}{|c|c|c|c|}\hline
direction in phase space               &$\epsilon_{y0}$=$\epsilon_{x0}$ &$\epsilon_{y0}$=0 &$\epsilon_{x0}$=0           \\ \hline
plateau level, $j_{trns}$              & 15e6,1.5e5,2e4 &8e6,1.5e6,1e4& 1.8e7,1.7e6,4e3       \\
plateau level,  $1/Log(j_{trns})$        &.139, .193, .232 &.145, .162 &.138,.160,.277         \\
platea level  seperation in $1/Log(j_{trns})$ & .054, .039   &.017& .022 \\ \hline
average plateau seperation     &.033 & &         \\
in $1/Log(j_{trns})$ &&&   \\   \hline
plateau width, $\Delta X_0$ (mm)       & 1.70, 1.83, 2.38 &2.0, 1.8 &2.2+, 2.0         \\ 
plateau width, $\Delta X_0/X_0$        &.075,.051,.106 &.095,.086 &.092, .083            \\  \hline
average plateau width, $\Delta X_0$ & 2.00 mm &    \\     \hline
\end{tabular}
\caption{Plateau parameters for three directions in phase space. $X_0=(\beta_{x0} (\epsilon_{x0}+\epsilon_{y0} ))^{.5}$.}
\label{tab.5-1}
\end{table}

The extrapolation depends on two parameters, the width of the plateaus and the
seperation between consecutive plateau levels as measured in terms of
$1/Log(j_{trns})$. The behaviour of these two parameters was studied by doing 
tracking runs for 3 different cases, corresponding to the 
three different directions in phase space,
$\epsilon_{y0}$=0, $\epsilon_{x0}$=$\epsilon_{y0}$ and $\epsilon_{x0}$=0. The two parameters were measured for these 
3 cases using runs of about 2e7 turns. The results are summarized in 
Table~\ref{tab.5-1}. 
Altogether, 7 plateaus were found and the two parameters 
for these 7 plateaus were measured. In Table~\ref{tab.5-1} one sees that the plateau width,
as measured as $\Delta X_0$, $X_0=(\beta_{x0} (\epsilon_{x0}+\epsilon_{y0} ))^{.5}$, is relatively constant 
with an average value of $\Delta X_0$=2.00 mm.
The plateau level seperation when measured in $1/Log(j_{trns})$ varies considerably
with an average value of .033.

	Based on the above results, it is proposed that in extrapolating the
survival function, it is assumed that the extrapolated plateau widths 
are  given by
$\Delta X_0$=2.00 mm, and the plateau level seperations in $1/Log(j_{trns})$ are .033.

\chapter{ Conclusions }

     Tracking studies lead  to a model of the survival function, which
pictures it as sequence of plateaus. Within the plateaus, the survival time in turns, $j_{trns}$, oscillates about 
some constant vaue of $j_{trns}$ which will be called the level of the plateau.
Studying the survival function along different directions in phase space, using an older version of the RHIC lattice , one finds that the width of the plateaus, 
$\Delta X_0$, $X_0=(\beta_{x0} (\epsilon_{x0}+\epsilon_{y0} ))^{.5}$, remains 
roughly constant at about 2.00 mm. The seperation between the levels of adjacent
plateaus has the same order of magnitude when measured in terms of the change in
$1/Log(j_{trns})$,  and has an average vaue of .033. Using, these results for the 
width of the plateaus and the seperation between plateau levels, one can 
extrapolate to estimate the location of the plateaus that correspond to longer
survival times than can be found by tracking. For the case treated, it was found
that a required survival time of 1e9 turns reduced the aperture by about 15\%
as compared to the aperture found by tracking using 1e6 turns.

The plateau model also leads to new criteria to be  used in tracking studies 
to find the aperture for  particles to  survive a given number of turns.
In the plateau model, one finds the first plateau whose level is higher than
the given number of turns, in order to find the aperture for the given number of turns.
This is to be compared with often used method, where one does a search starting at
large amplitudes until one finds an amplitude that survives the given number
of turns. In the latter method one cannot be sure that a finer search would not 
find unstable runs at smaller amplitudes or how frequently these unstable runs
will occur. In the plateau model, while there may be unstable runs at smaller amplitudes, there is the assumption that they will not occur frequently.

\end{document}